\documentclass[journal,transmag]{IEEEtran}
\usepackage{amsmath}
\usepackage{amssymb}
\usepackage{mathtools}
\usepackage{hyperref,color,breqn,multirow}
\usepackage[top=0.7in, left=0.65in]{geometry}
\usepackage{graphicx}
\usepackage{todonotes}
\usepackage{tikz}
\usepackage{amsthm}
\usepackage[utf8]{inputenc}
\usepackage{eso-pic}

\newtheorem*{remark}{Remark}
\usepackage[siunitx,europeanresistors,americaninductors]{circuitikz}
\usetikzlibrary{external}
\newcommand{\IKR}[1]{{\textcolor{black}{#1}}}

\AddToShipoutPicture*{\footnotesize\sffamily\raisebox{1cm}{\hspace{1.65cm}\fbox{\parbox{\textwidth}{\copyright~2020 IEEE. Personal use of this material is permitted. Permission from IEEE must be obtained for all other uses, in any current or future media, including reprinting/republishing this material for advertising or promotional purposes, creating new collective works, for resale or redistribution to servers or lists, or reuse of any copyrighted component of this work in other works.}}}}

\begin{document}
\title{Efficient Simulation of Field/Circuit Coupled Systems\\ with Parallelised Waveform Relaxation}
\author{\IEEEauthorblockN{Idoia Cortes Garcia\IEEEauthorrefmark{1,2},
Iryna Kulchytska-Ruchka\IEEEauthorrefmark{1,2}, and
Sebastian Sch\"ops\IEEEauthorrefmark{1,2}}\thanks{Manuscript received xxx y, 20zz; revised xxx yy, 20zz and xxx 1, 20zz; accepted xxx 1, 20zz. Date of publication xxx yy, 20zz; date of current version xxx yy, 20zz. (Dates will be inserted by IEEE; published is the date the accepted preprint is posted on IEEE Xplore; current version is the date the typeset version is posted on Xplore). Corresponding author: F. A. Author (e-mail: f.author@nist.gov). Digital Object Identifier (inserted by IEEE).}
\IEEEauthorblockA{\IEEEauthorrefmark{1}Institut f\"ur Teilchenbeschleunigung und Elektromagnetische Felder (TEMF), Technische Universit\"at Darmstadt, Germany}
\IEEEauthorblockA{\IEEEauthorrefmark{2}Centre for Computational Engineering (CCE), Technische Universit\"at Darmstadt, Germany}}

\IEEEtitleabstractindextext{\begin{abstract}
This paper proposes an efficient parallelised computation of field/circuit coupled systems co-simulated with the Waveform Relaxation (WR) technique.
{The main idea of the introduced approach lies in application of the parallel-in-time method parareal to the WR framework. 
Acceleration obtained by the time-parallelisation is further increased in the context of micro/macro parareal. Here, the field system is replaced by a lumped model in the circuit environment for the sequential computations of parareal.} 
The {introduced} algorithm is tested with {a model} of a {single-phase isolation} transformer coupled to a rectifier circuit.
\end{abstract}

}

\maketitle
\thispagestyle{empty}
\pagestyle{empty} 

\section{Introduction}
\begin{figure}[t]
	\centering
	\scalebox{0.8}{
		\includegraphics{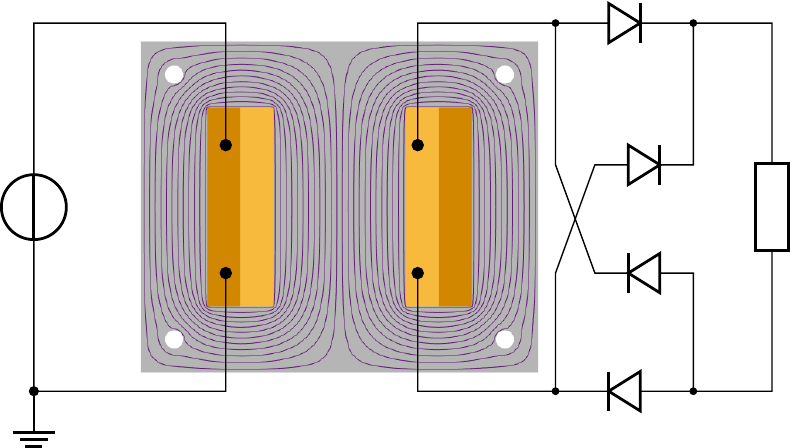}
	}
	\vspace{-0.3em}
	\caption{Transformer model `MyTransformer' coupled to a rectifier circuit \cite[Section 6.3]{Schops_2011ac} and \cite{Meeker_2018aa}, see \url{http://www.femm.info/wiki/MyTransformer}.}
	\label{fig:Transformer}
	\vspace{-1em}
\end{figure}

Simulation of devices and their surrounding circuitry is often performed with circuit simulators. {Within such simulations,} the behaviour of the devices is described by lumped element models, which yield algebraic  or differential relations between the voltages and {the} currents. A disadvantage of the lumped models is that they do not provide enough details, 
whenever a spatial description of the electromagnetic field inside a device is needed. 
Such cases include for example the simulation of electric machines \cite{Salon_1995aa} or of the quench protection system of superconducting magnets in particle accelerators \cite{Bortot_2018ab}. In these cases field/circuit coupling \cite{Bedrosian_1993aa, Schops_2011ac, Cortes-Garcia_2017ab} is needed, see Fig.~\ref{fig:Transformer}. In order to exploit the different time rates {and also to be} able to use separate dedicated solvers for the different systems of equations involved, waveform relaxation (WR) is often used~\cite{Lelarasmee_1982ab}. Here, the different systems of equations are solved separately and, iteratively, information is exchanged between them until they converge to the coupled solution.

For the numerical solution of the field system, space  is discretised first e.g. with the finite element method (FEM). Together with the circuit equations this results in systems of differential algebraic equations (DAEs), which have to be solved in the time domain. The finer the mesh, the larger are the systems to be solved at every time step. This leads to long computational time. To this end, calculations can be accelerated by means of a parallel-in-time method called parareal, which is a specific shooting method \cite{Lions_2001aa, Gander_2015aa}.

This work combines parareal and WR together in one algorithm. In contrast to previous works, e.g. \cite{Liu_2012aa, Cadeau_2011aa}, engineering knowledge is used to design a new and optimised algorithm which significantly reduces the computational cost borrowing ideas from micro/macro parareal \cite{Maday_2007aa,Legoll_2013aa}.

The structure of the paper is the following: Section 2 introduces the spatially discretised systems of equations. Section 3 explains the waveform relaxation algorithm with an optimised transmission condition for the field/circuit coupled case. In section 4 the classical as well as a special case of the micro/macro parareal algorithm are introduced. Section 5 deals with the coupling of waveform relaxation and parareal and finally  section 6 presents numerical simulations. The last section closes the paper with conclusions and an outlook to the future work.

\section{Systems of equations}
To describe the electromagnetic field part, we consider a magnetoquasistatic approximation of Maxwell's equations in terms of the reduced A formulation \cite{Emson_1988aa}.  This leads to the curl-curl eddy current partial differential equation (PDE) that describes the field in terms of a magnetic vector potential. 
The circuit side is  formulated with the modified nodal analysis (MNA) \cite{Ho_1975aa}. For the numerical simulation of the coupled system, the method of lines is used. This leads to a time-dependent coupled system of DAEs that are formulated as an initial value problem (IVP). 

For $t\in\mathcal{I} = (T_0,T_N]$ we solve the IVP described by the coupled system. The field DAEs are
\begin{align}
		{\mathbf{M}}\frac{\mathrm{d}}{\mathrm{d}t}{\mathbf{a}} + \mathbf{K}\big(\mathbf{a}\big)\mathbf{a} &= \mathbf{X}\mathbf{i}_{\mathrm{m}},\label{eq:sys1}\\
	\mathbf{X}^{\top}\frac{\mathrm{d}}{\mathrm{d}t}\mathbf{a}& = \mathbf{v}_{\mathrm{m}},
\end{align}
where $\mathbf{a}$ is the discretised magnetic vector potential, $\mathbf{v}_{\mathrm{m}}$ the voltage across and $\mathbf{i}_{\mathrm{m}}$ the current through the coil of the electromagnetic device. $\mathbf{M}$ denotes the (singular) mass matrix, $\mathbf{K}$ the (possibly gauged) curl-curl matrix, and $\mathbf{X}$ distributes the circuit's input currents in space based on the {stranded conductor model}. 
In two dimension, we consider either 
$$\mathbf{M}_{i,j}=\int_{\Omega}\alpha_i\cdot\sigma\alpha_j\text{~~or~~} \mathbf{M}_{i,j}=\int_{\Omega}\frac{1}{12}d^2\nabla\alpha_i\sigma\nabla\alpha_j$$if lamination is considered, \cite{Gyselinck_1999aa}, where $\alpha_{\star}$ are test and weighting functions from an appropriate space defined on the computational domain $\Omega$, $\sigma$ the conductivity and $d$ the lamination thickness. 
We assume that the spaces contain boundary and gauging condition (e.g. tree/cotree) if necessary.
The circuit is described in terms of the MNA by the system
\begin{align}
\mathbf{A}\frac{\mathrm{d}}{\mathrm{d}t}{\mathbf{x}} + \mathbf{B}\big(\mathbf{x}\big)\mathbf{x} + \mathbf{P}\mathbf{i}_{\mathrm{c}} = \mathbf{f}(t),\\
\mathbf{P}^{\top}\mathbf{x} = \mathbf{v}_{\mathrm{c}},\label{eq:sys4}
\end{align}
with $\mathbf{x}$ containing the node potentials and currents through branches with voltage sources and inductors, $\mathbf{v}_{\mathrm{c}}$ the voltage across and $\mathbf{i}_{\mathrm{c}}$ the current through branches containing the electromagnetic element, $\mathbf{A}$ and $\mathbf{B}$ being the MNA system matrices and $\mathbf{P}$  the incidence matrix of the field element describing its position inside the circuit's graph. 
For the coupling, $\mathbf{v}_{\mathrm{c}} = \mathbf{v}_{\mathrm{m}}$ and $\mathbf{i}_{\mathrm{c}} = \mathbf{i}_{\mathrm{m}}$ and
given initial values 
\begin{align}\label{eq:IC}
\mathbf{a}(t_0) = \mathbf{a}_0,\; \mathbf{x}(t_0)=\mathbf{x}_0,\; \mathbf{i}_{\mathrm{c}}(t_0)=\mathbf{i}_0\; \text{and} \;\mathbf{v}_{\mathrm{c}}(t_0) = \mathbf{v}_0,
\end{align}
the coupled system \eqref{eq:sys1}-\eqref{eq:sys4} can be solved in time.

\section{Waveform Relaxation}
We start the WR algorithm by dividing the simulation time span $\mathcal{I}$  into $N$ time windows $\mathcal{I}_{n}=(T_{n-1}, T_{n}]$ of size $\Delta T$. At iteration $k+1$, a Gauss-Seidel scheme is applied to \eqref{eq:sys1}-\eqref{eq:sys4}, which allows to solve the field and circuit systems of equations separately and iteratively exchange information between them until the solution converges up to a certain tolerance. 

For each time window $\mathcal{I}_{n}$ and WR iteration $k+1$, the algorithm starts by solving the field system
\begin{align}
	{\mathbf{M}}\frac{\mathrm{d}}{\mathrm{d}t}{\mathbf{a}}^{(k+1)} + \mathbf{K}\big(\mathbf{a}^{(k+1)}\big)\mathbf{a}^{(k+1)} &= \mathbf{X}\mathbf{i}^{(k+1)}_{\mathrm{m}},\label{eq:1}\\
	\mathbf{X}^{\top}\frac{\mathrm{d}}{\mathrm{d}t}\mathbf{a}^{(k+1)}& = \mathbf{v}^{(k+1)}_{\mathrm{m}},\\
\mathbf{v}^{(k+1)}_{\mathrm{m}} &= \mathbf{v}^{(k)}_{\mathrm{c}}.\label{eq:TCfield}
\end{align}
In the first WR iteration, $\mathbf{v}^{(0)}_{\mathrm{c}}$ is computed by (constant) extrapolation of the initial condition $\mathbf{v}^{(0)}_{\mathrm{c}}(T_{n-1})$.
Afterwards, the circuit can be solved independently with
\begin{align}
	\mathbf{A}\frac{\mathrm{d}}{\mathrm{d}t}{\mathbf{x}}^{(k+1)} + \mathbf{B}\big(\mathbf{x}^{(k+1)}\big)\mathbf{x}^{(k+1)} + \mathbf{P}\mathbf{i}^{(k+1)}_{\mathrm{c}} = \mathbf{f}(t) \label{eq:circ1},\\
	\mathbf{P}^{\top}\mathbf{x}^{(k+1)} = \mathbf{v}^{(k+1)}_{\mathrm{c}} \label{eq:circ2},\\
		\mathbf{v}_{\mathrm{c}}^{(k+1)} = \mathbf{L}^{(k+1)}\frac{\mathrm{d}}{\mathrm{d}t}\mathbf{i}_{\mathrm{c}}^{(k+1)} -\mathbf{L}^{(k+1)}\frac{\mathrm{d}}{\mathrm{d}t}\mathbf{i}_{\mathrm{m}}^{(k+1)} + \mathbf{v}_{\mathrm{m}}^{(k+1)}\label{eq:TCopt},
\end{align}
where the inductance
\begin{equation}\label{eq:mutualinductance}
	\mathbf{L}^{(k+1)} = \mathbf{X}^{\top}\mathbf{K}^+(\mathbf{a}^{(k+1)})\mathbf{X}
\end{equation}
is used to optimise convergence \cite{Schops_2011ac, Cortes-Garcia_2017ab}. Here, $\mathbf{K}^+$ denotes the pseudo-inverse, if necessary due to gauging.
These steps are repeated until the difference of the solutions between two subsequent iterations is small enough. Once convergence up to a certain tolerance is reached, the solution obtained at time $T_n$ is used as initial condition for the next time window $\mathcal{I}_{n+1}$ and the iteration scheme can be repeated again. Thus, the WR algorithm is performed sequentially through all time windows.

Considering the field system's transmission condition \eqref{eq:TCfield}, the analogous version for the circuit part would be to set
$$\mathbf{i}_{\mathrm{c}}^{(k+1)} = \mathbf{i}_{\mathrm{m}}^{(k+1)}.$$
\IKR{However, instead, \eqref{eq:TCopt} is used. This yields an improved exchange of information between the two systems in the context of optimised Schwarz methods \cite{Al-Khaleel_2014aa}. It corresponds to  treating the field as an inductor with a correction voltage source. It has been shown that this leads to faster WR convergence \cite{Schops_2011ac, Cortes-Garcia_2017ab}.}

\section{Parareal}
The parareal algorithm starts with partitioning the interval $\mathcal{I}$ into (the same) windows $\mathcal{I}_n=(T_{n-1}, T_{n}]$ of size $\Delta T$. 
We start by summarizing \eqref{eq:sys1}-\eqref{eq:IC} as the initial-value problem
\begin{equation}
\mathbf{C}\frac{\mathrm{d}}{\mathrm{d}t}\mathbf{u}=\mathbf{g}(t,\mathbf{u}),\quad\mathbf{u}(T_0)=\mathbf{u}_0,\label{eq:fine}
\end{equation}
with $\mathbf{u}:\mathcal{I}\rightarrow\mathbb{R}^{n_{\mathrm{dof}}}$ and $n_{\mathrm{dof}}$ the degrees of freedom (DoF) of the coupled system. Within the parareal framework, one solves
\begin{equation}
\mathbf{C}\frac{\mathrm{d}}{\mathrm{d}t}\mathbf{u}_n = \mathbf{g}(t, \mathbf{u}_n), 
\quad{\mathbf{u}_n(T_{n-1})= \mathbf{U}_{n-1}},  
\quad{t} {\in \mathcal{I}_n},\label{eq:pbm_subint}
\end{equation}
on each subinterval $\mathcal{I}_n$ in parallel, starting from an initial value $\mathbf{U}_{n-1},$ $n=1,\dots,N$ with a given $\mathbf{U}_0:=\mathbf{u}_0.$ 
The goal is then to eliminate the mismatch between the values at synchronisation points $T_n,$ $n=1,\dots,N-1.$ 
The parareal iteration reads \cite{Lions_2001aa}: for $k=0,1,\dots,K$ and $n=1,\dots,N$
\begin{align}
{{\mathbf{U}_0^{(k+1)}}}&
=\mathbf{u}_0,\nonumber\\
{{\mathbf{U}_n^{(k+1)}}}&
=
{\mathcal{F}}\big(T_n, T_{n-1},{\mathbf{U}^{(k)}_{n-1}}\big)\label{eq:Parareal}\\
&\quad+{{\mathcal{G}}}\big(T_n, T_{n-1},{\mathbf{U}^{(k+1)}_{n-1}}\big) - {{\mathcal{G}}}\big(T_n, T_{n-1},{\mathbf{U}^{(k)}_{n-1}}\big).\nonumber
\end{align}
\noindent
Operators ${\mathcal{F}}$ and ${\mathcal{G}}$ in  \eqref{eq:Parareal} are a fine and a coarse propagator, respectively.
${\mathcal{F}}\big(T_n, T_{n-1},\mathbf{U}^{(k)}_{n-1}\big)$ gives an accurate solution (e.g. using small time steps) of \eqref{eq:pbm_subint}, starting from initial values $\mathbf{U}^{(k)}_{n-1}$ and can be calculated in parallel on all $\mathcal{I}_{n}$. On the other hand, the coarse solution ${{\mathcal{G}}}\big(T_n, T_{n-1},\mathbf{U}^{(k+1)}_{n-1}\big)$ is obtained in a cheaper way (e.g. using large time steps) but sequentially.

\subsection{Micro/Macro Parareal}
The idea of micro/macro parareal is to consider different models on the levels. On the fine level, the original problem \eqref{eq:fine} is solved. However, for the coarse one, a reduced IVP
\begin{equation}
\mathbf{C}_{\mathrm{r}}\frac{\mathrm{d}}{\mathrm{d}t}\mathbf{u}_{\mathrm{r}}=\mathbf{g}_{\mathrm{r}}(t,\mathbf{u}_{\mathrm{r}}),\quad\mathbf{u}_{\mathrm{r}}(T_0)=\mathbf{u}_{\mathrm{r},0}\label{eq:coarse}
\end{equation}
is considered, with $\mathbf{u}_{\mathrm{r}}:\mathcal{I}\rightarrow\mathbb{R}^{n_{\mathrm{r}}}$ and $n_{\mathrm{r}}$ the number of DoFs of the reduced system.  This allows to further speed up the computation of the coarse solution by not only saving the cost of the time stepper, but also reducing the size of the system solved. Here, for example, model order reduction techniques  can be used \cite{Maday_2007aa} to set up the simplified coarse system in \eqref{eq:coarse}.

Now, two additional operators have to be defined, in order to exchange the solutions between the models. The restriction operator $\mathcal{R}:\mathbb{R}^{n_{\mathrm{dof}}}\rightarrow\mathbb{R}^{n_{\mathrm{r}}}$ allows to, given an initial value of \eqref{eq:fine}, obtain a valid initial value for \eqref{eq:coarse}. The inverse can be done with the lifting operator $\mathcal{L}:\mathbb{R}^{n_{\mathrm{r}}}\rightarrow\mathbb{R}^{n_{\mathrm{dof}}}$, which, given an initial value of the coarse system, computes a valid initial value for the fine one. They are defined consistently, such that
\begin{equation}
	\mathcal{R}(\mathcal{L}(X)) = X.\label{eq:RL}
\end{equation} 
With these operators, the update in \eqref{eq:Parareal} is changed to \cite{Maday_2007aa}
\begin{align}
	\label{eq:MMParareal}
	{{\mathbf{U}_n^{(k+1)}}}&
	=
	\tilde{\mathbf{U}}_n^{(k)}
	+\mathcal{L}\left(\bar{\mathbf{U}}_n^{(k+1)}\right) 
	-\mathcal{L}\left(\bar{\mathbf{U}}_n^{(k)}\right),\end{align}
with fine $\tilde{\mathbf{U}}_n^{(k)} \coloneqq {\mathcal{F}}\big(T_n, T_{n-1},{\mathbf{U}^{(k)}_{n-1}}\big)$ and coarse solution $\bar{\mathbf{U}}_n^{(k+1)} \coloneqq {{\mathcal{G}}}\big(T_n, T_{n-1},{\mathcal{R}(\mathbf{U}^{(k+1)}_{n-1})}\big)$.

\section{Parallelised Waveform Relaxation}
For the coupling of WR with parareal (PRWR), we consider the WR and the parareal window sizes to be the same $\Delta T$. This choice is not necessary, however a natural one. We present two variants of the algorithm.

The initial version of the algorithm is similar to \cite{Cadeau_2011aa}. Here, we {choose} both the coarse {and} the fine propagator{s to be a} WR scheme. For the the coarse propagator, the WR scheme will not iterate until convergence, but will stop after a finite (fixed) number of iterations $k_{\mathrm{c}}$. In contrast to  \cite{Cadeau_2011aa}, the fine propagator iterates the WR scheme until convergence up to a certain tolerance.

\begin{figure}[t]
	\centering
	\scalebox{0.85}{
		\includegraphics{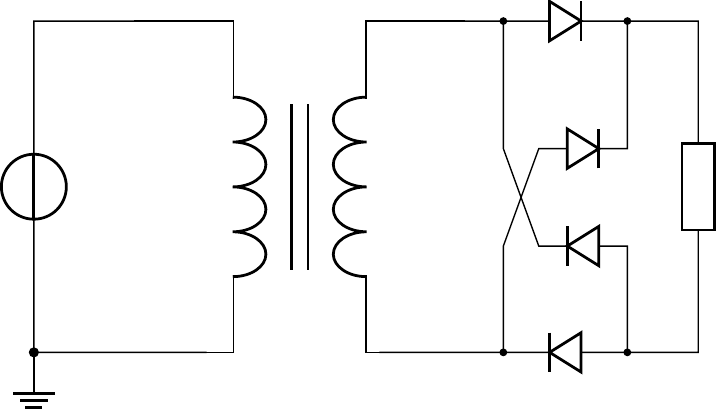}
	}
	\caption{Micro/macro parareal coarse system model.}
	\label{fig:Coarsetrans}
\end{figure}

The second algorithm is an optimised version of the first one, where the coarse propagator is chosen to only perform half of a WR iteration ($k_{\mathrm{c}}=0.5$) by starting with the circuit system. This means only solving \eqref{eq:circ1}-\eqref{eq:TCopt} for $t\in\mathcal{I}$ and replacing the transformer by a mutual inductor as depicted in Fig.~\ref{fig:Coarsetrans} extracted by \eqref{eq:mutualinductance}. It can also be interpreted as neglecting the eddy current effects of the magnetoquasistatic problem on the coarse level. This allows to significantly reduce the computational cost as the coarse propagator only operates on the circuit level, i.e. with rather few DoFs. This algorithm fits into the context of micro/macro parareal algorithms \cite{Maday_2007aa,Legoll_2013aa}.
 
\section{Numerical simulations}
We apply the \IKR{two} introduced approaches to a 2D model of a single-phase isolation transformer (`MyTransformer') coupled to a rectifier circuit \cite[Fig.~6.6~(b)]{Schops_2011ac}, which we depict in Fig.~\ref{fig:Transformer}. For the numerical simulations we consider field-independent materials, such that the eddy-current curl-curl equation \eqref{eq:sys1} is a linear DAE. 
The simulation interval is chosen as $\mathcal{I}=[0, 0.1]$ s, which is partitioned into $N$ windows. For the time integration inside the WR scheme of both coarse and fine propagators, implicit Euler method is used. The fine time step size is $h=5\cdot10^{-5}$ s and the excitation voltage source is $v(t)=220\sin(400\pi t)$ V. Parareal is performed until the relative $l^2$ error of the jumps is below $10^{-5}$. 

Firstly the initial algorithm is applied such that the coarse propagator performs $k_{\mathrm{c}}=1.5$ WR iterations with time step size $\Delta T$ for both subsystems. The fine propagator executes WR until convergence (relative $l^2$ error of the coupling variables between two subsequent iterations $<10^{-8}$).

In the optimised case, \IKR{the coarse solver only solves the circuit part with a lumped inductance element replacing the field system. This system is solved with a time step size of  $\Delta T$.  For the fine propagator, the solution of the field/circuit coupled problem with WR until convergence (again relative error $<10^{-8}$) is used.} 	
In this case, a micro/macro parareal algorithm arises. Therefore, restriction and lifting operators must be constructed. The DoFs of the fine system $\mathbf{u}_{\mathrm{f}}$ are
$$\mathbf{u}_{\mathrm{f}}^{\top} = (\mathbf{a}^{\top}, \mathbf{i}_{\mathrm{m}}^{\top}, \mathbf{v}_{\mathrm{m}}^{\top}, \mathbf{x}^{\top}, \mathbf{i}_{\mathrm{c}}^{\top}, \mathbf{v}_{\mathrm{c}}^{\top}),$$
whereas the coarse operator reduces them to $\mathbf{u}_{\mathrm{r}}=\mathbf{u}_{\mathrm{c}}$, with
$$\mathbf{u}_{\mathrm{c}}^{\top} = (\mathbf{x}^{\top}, \mathbf{i}_{\mathrm{c}}^{\top}, \mathbf{v}_{\mathrm{c}}^{\top}).$$
We set the restriction  operator $\mathcal{R}$ to
$$\mathcal{R}(\mathbf{u}_{\mathrm{f}}) =(\mathbf{x}^{\top}, \mathbf{i}_{\mathrm{c}}^{\top}, \mathbf{v}_{\mathrm{c}}^{\top})^{\top}.$$
The lifting $\mathcal{L}$ computes the corresponding magnetic vector potential $\mathbf{a}$ obtained from solving a magnetostatic problem with a given current, that is,
$$\mathcal{L}(\mathbf{u}_{\mathrm{c}}) = ((\mathbf{K}^{+}\mathbf{X}\mathbf{i}_{\mathrm{c}})^{\top}, \mathbf{i}_{\mathrm{c}}^{\top}, \mathbf{v}_{\mathrm{c}}^{\top}, \mathbf{x}^{\top}, \mathbf{i}_{\mathrm{c}}^{\top}, \mathbf{v}_{\mathrm{c}}^{\top})^{\top}.$$
Please note that, as stated in \eqref{eq:RL}, $\mathcal{R}\big(\mathcal{L}(\mathbf{u}_{\mathrm{c}})\big) = \mathbf{u}_{\mathrm{c}}.$
\begin{remark}
	In practice, for nonlinear materials, in each parareal iteration and window, the inductance \eqref{eq:mutualinductance} is extracted at a working point and kept constant. 
\end{remark}

\subsection{Simulation results}
\begin{figure}[t]
	\centering
	\includegraphics{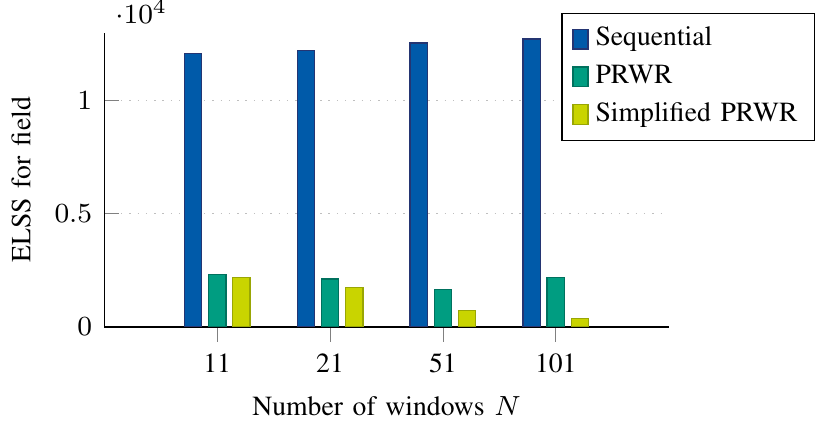}
		
	\includegraphics{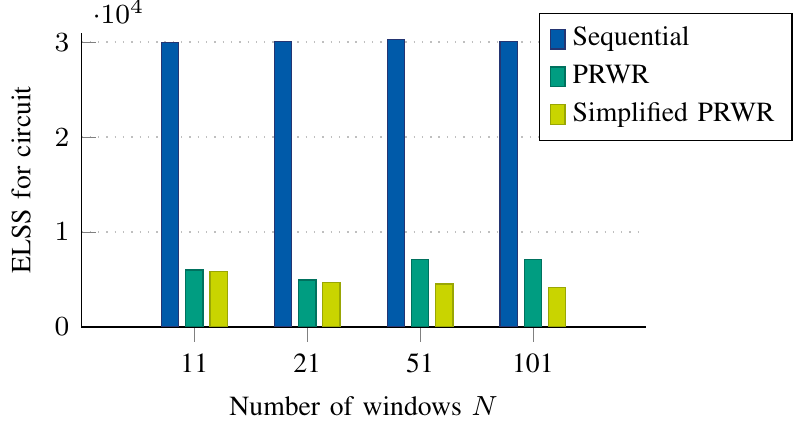}
		\vspace{-0.6em}
	\caption{Effective number of linear system solves (ELSS), i.e. neglecting the ones that can be carried out in parallel.}
	\label{fig:improvement}
\end{figure}

Both algorithms are applied to the same transformer with the eddy current lamination model \cite{Gyselinck_1999aa}. The effective number of linear system solves for sequential WR, PRWR with the first algorithm and PRWR with the second algorithm is shown in Fig.~\ref{fig:improvement}. It can be seen that, even though both PRWR algorithms significantly decrease the effective number of solutions of linear systems compared to the sequential simulation, the first algorithm is sooner affected by the sequential computation of the coarse system, which increases the effective cost for larger number of processors~$N$. The optimised algorithm profits here from the fact that no FEM simulations are necessary on the coarse level.

\section{Future work}
The simplified algorithm has also been applied to the eddy current  problem without lamination of the core. However, the algorithm only converges after the $N$th parareal iteration, which 
does not yield a speed up with respect to the sequential simulation. This is due to the fact that the eddy current effects are dominant and the static simplification made 
in the coarse system is not accurate enough. There are two main sources of error arising from the approximation. First, neglecting the eddy current losses leads to a different current to voltage relation on the circuit side. A second source of error is the lifting operator, as it distributes the magnetic vector potential neglecting the skin effect which, for the considered example, is highly relevant. Future work will investigate more sophisticated lumped models and liftings.

\section{Conclusion}
This paper proposes a parallel-in-time framework for efficient solution of field/circuit coupled problems. Two approaches of a joint application of the co-simulation technique and parareal are introduced and applied to a single-phase isolation transformer. The results confirm that both approaches converge quickly and allow to reduce the effective computational cost compared to the sequential waveform relaxation simulation. In the best case the time spent on FE computations is sped up by a factor of 33
, i.e. for $N=101$ comparing the field
ELSS of Sequential and Simplified PRWR (see Fig.~\ref{fig:improvement}). 

\vspace{-0.05cm}
\section*{\small Acknowledgement}
{\footnotesize This work is supported by the `Excellence Initiative' of the German Federal and State Governments
and the Graduate School of Computational Engineering at TU Darmstadt and DFG Grant SCHO1562/1-2 and BMBF Grant 05M2018RDA (PASIROM). \IKR{The authors would also like to thank Lorenzo Bortot and Micha\l{} Maciejewski for the fruitful discussions.}\par}

\begin{figure}[t]
	\centering
	\includegraphics[height=3cm]{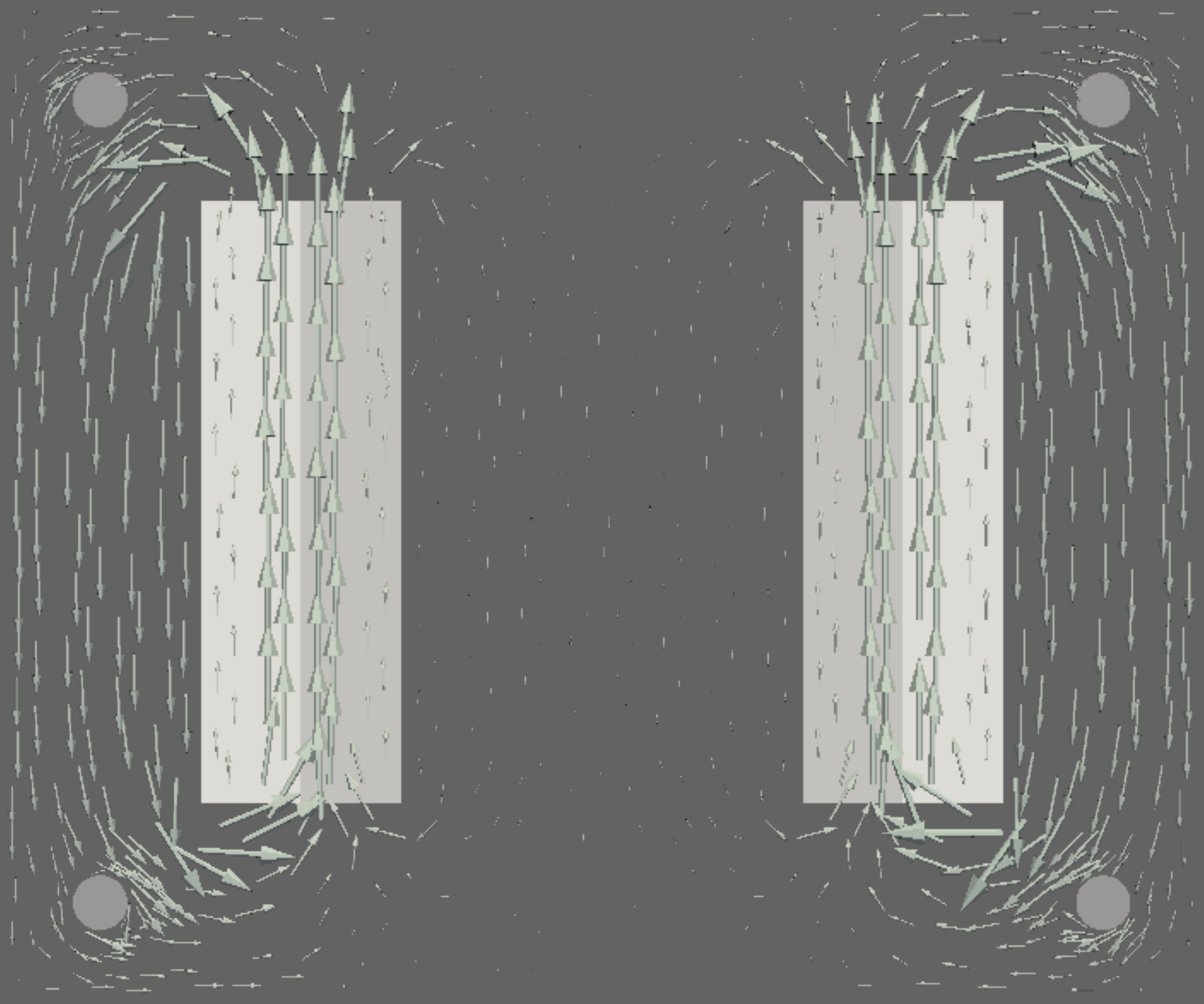}~
	\includegraphics[height=3cm]{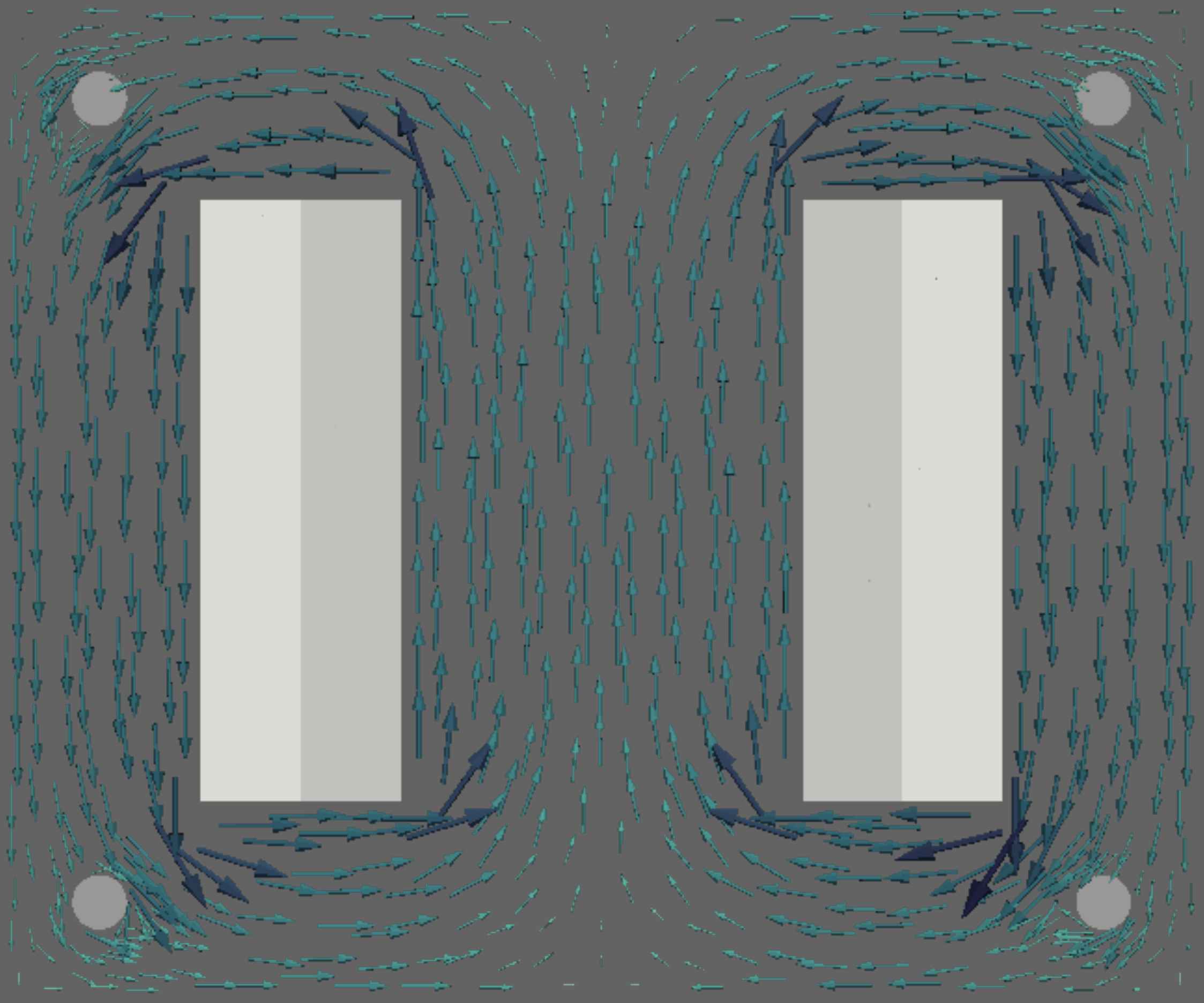}~
	\includegraphics[height=3cm]{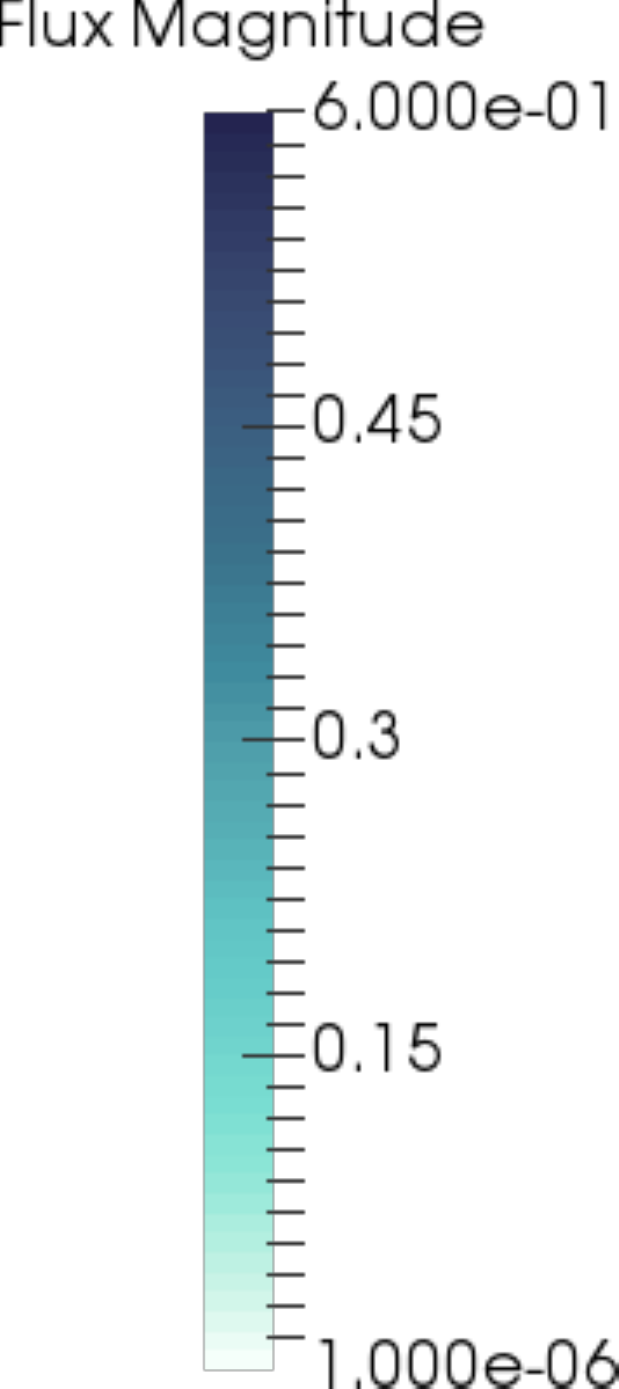}
	\caption{Coarse $\mathcal{L}(\bar{\mathbf{U}}_n^{(1)})$ (left) and fine field solutions $\tilde{\mathbf{U}}_n^{(3)}$ (right), $n=100$ and $N=101$.}
	\vspace{-1em}
\end{figure}

\end{document}